\documentclass[11pt,a4paper,english,superscriptaddress,prd,aps,preprint]{revtex4}
\usepackage{amssymb}
\usepackage{amsmath} 
\usepackage{slashed}
\usepackage{graphicx}
\usepackage{babel}
\usepackage{hyperref}
\usepackage{color}
\makeatletter
\AtBeginDocument{\let\@elt\relax}\makeatother

\begin{document}

\title{Gravitational corrections to two-loop beta function in quantum electrodynamics}

\author{L.~Ibiapina~Bevilaqua}
\email{leandro@ect.ufrn.br}
\affiliation{Escola de Ci\^encias e Tecnologia, Universidade Federal do Rio Grande do Norte\\
Caixa Postal 1524, 59072-970, Natal, Rio Grande do Norte, Brazil.}

\author{M. Dias}
\email{marco.dias@unifesp.br}
\affiliation{Universidade Federal de S\~ao Paulo,
Departamento de F\'isica, Rua S\~ao Nicolau 210,
09913-030, Diadema, S\~ao Paulo, Brazil.}

\author{A.~C.~Lehum}
\email{lehum@ufpa.br}
\affiliation{Faculdade de F\'isica, Universidade Federal do Par\'a, 66075-110, Bel\'em, Par\'a, Brazil.}

\author{C. R. Senise Jr.}
\email{carlos.senise@unifesp.br}
\affiliation{Universidade Federal de S\~ao Paulo,
Departamento de F\'isica, Rua S\~ao Nicolau 210,
09913-030, Diadema, S\~ao Paulo, Brazil.}

\author{A. J. da Silva}
\email{ajsilva@fma.if.usp.br}
\affiliation{Instituto de F\'\i sica, Universidade de S\~ao Paulo\\
Caixa Postal 66318, 05315-970, S\~ao Paulo, S\~ao Paulo, Brazil.}

\author{Huan Souza}
\email{huan.souza@icen.ufpa.br}
\affiliation{Faculdade de F\'isica, Universidade Federal do Par\'a, 66075-110, Bel\'em, Par\'a, Brazil.}

\begin{abstract}
In this work, we use the framework of effective field theory to couple Einstein's gravity to quantum electrodynamics (QED) and determine the gravitational corrections to the two-loop beta function of the electric charge in arbitrary electrodynamics (Lorentz-like) and arbitrary (de Donder like) gravitational gauges. Our results indicate that gravitational corrections do not alter the running behavior of the electric charge; on the contrary, we observe that it gives a positive contribution to the beta function, making the electric charge grow even faster.
\end{abstract}


\maketitle

\section{Introduction}

Even though the classical theory of the gravitational field is well established by Einstein's theory of general relativity, the quantum description of gravity is still an open problem. Several approaches concur to describe gravity at the quantum level. As nonperturbative approaches we can remark asymptotic aafety~\cite{Percacci:2007sz}, causal dynamical triangulations~\cite{Loll:2019rdj}, lattice quantum gravity~\cite{Hamber:2009mt}, and loop quantum gravity~\cite{Rovelli:1997yv}, besides superstrings as a possible unifying theory of all forces. On the other side, perturbative quantization of Einstein's theory of gravity, for small fluctuations, around a flat metric leads to a nonrenormalizable quantum field theory~\cite{'tHooft:1974bx,PhysRevLett.32.245,Deser:1974cy}. 

However, the potential harm of nonrenormalizability in the perturbative approach can be contoured in the effective field theory (EFT) framework, where there is an unambiguous way to define a well-behaved and reliable quantum theory of gravitation, if only we agree to restrict to processes of low energy compared to the Planck scale~\cite{Donoghue:2017pgk,Burgess:2003}. This EFT approach has been applied to compute quantum gravitational corrections to several quantities such as Newtonian and Coulomb potentials \cite{Donoghue:1994dn,Faller:2007sy,Donoghue:2012zc}, the Friedmann-Lema\^itre-Robertson-Walker metric~\cite{Donoghue:2014yha}, bending of light by a massive source \cite{Bjerrum-Bohr:2014zsa}, and others.

Although the effective field theory of gravitation is perfectly well defined as a quantum field theory, some subtleties arise from its nonrenormalizability, such as the use of the renormalization group equations~\cite{Weinberg:1979, Buchler:2003vw}, as illustrated by the controversy involving the gravitational corrections to the beta function of gauge theories. In 2006, Robinson and Wilczek announced their conclusion that gravity contributes with a negative term to the beta function of the gauge coupling constant, opening the possibility that the coupling to quantum gravity could make gauge theories asymptotically free~\cite{Robinson:2005fj}. This result was soon contested in Ref.~\cite{Pietrykowski:2006xy}, where it was shown that the claimed gravitational correction is gauge dependent, and a lot of subsequent research on the subject followed with varying conclusions~\cite{Felipe:2012vq, Ebert:2007, Nielsen:2012fm, Toms:2008dq, Toms:2010vy, Toms:2011zza, Donoghue:2012zc, Ellis:2010, Anber:2010uj, Park:2018vci}.

In particular, it was shown through the computation of scattering processes at one loop, that quantum gravitational corrections do not alter the running behavior of the electric charge in the massless Einstein-Scalar quantum electrodynamics (QED)~\cite{Charneski:2013zja}, and the same behavior was observed in the massive case~\cite{Bevilaqua:2015hma}.

In this work, we couple gravity to Abelian gauge fields and matter, both fermionic and scalar, in order to compute the gravitational corrections to the beta function of the electric charge at two loops. We start with fermionic QED in Sec. \ref{sec1}, where we discuss the results for one- and two-loop (Secs. \ref{sec1-oneloop} and \ref{sec1-twoloop}) contributions. In Sec. \ref{sec2} we compute the one- and two-loop beta function of the electric charge in the scalar QED. In Sec. \ref{summary} we present some final remarks. 
 
Throughout this paper, we use natural units $c=\hbar=1$.

\section{Quantum electrodynamics coupled to gravity}\label{sec1}

The Lagrangian describing gravity coupled to QED is
\begin{eqnarray}\label{fQED}
\mathcal{L}=&& \sqrt{-g}\Big\{\frac{2}{\kappa^2}R-\frac14 g^{\mu\nu}g^{\alpha\beta} F_{\mu\alpha} F_{\nu\beta} +\frac{i}{2} \left[\bar{\psi}\gamma^{\mu}(\overrightarrow{\nabla}_{\mu} - ieA_{\mu})\psi-\bar{\psi}(\overleftarrow{\nabla}_{\mu} + ieA_{\mu})\gamma^{\mu}\psi\right] - m\bar{\psi}\psi\Big\} \nonumber \\
&&\qquad\qquad +\mathcal{L}_{HO} +\mathcal{L}_{GF}+\mathcal{L}_{CT},
\end{eqnarray}
\noindent where the Dirac matrices are contracted with the vierbein ($\gamma^{\mu} \equiv \gamma^a e^{\mu}_a$), $\overrightarrow{\nabla}_{\mu}\psi=(\partial_\mu +i\omega_\mu)\psi$, $\bar\psi\overleftarrow{\nabla}_{\mu}=(\partial_\mu\bar\psi -i\bar\psi\omega_\mu)$, $\omega_\mu=\frac{1}{4}\sigma^{ab}\left[e^\nu_a(\partial_\mu e_{b\nu}-\partial_\nu e_{b\mu})
+\frac{1}{2}  e^\rho_a e^\sigma_b(\partial_\sigma e_{c\rho}-\partial_\rho e_{c\sigma})e^c_\mu-(a\leftrightarrow b)\right]$ is the spin connection with $\sigma^{ab}=i[\gamma^a , \gamma^b]/2$, $e$ is the electric charge and $\kappa$ is the gravitational coupling ($\kappa^2=32\pi G=32\pi/M_P^2$), with $M_P$ being the Planck mass and $G$ the Newtonian gravitational constant. 

An observation is in order. Even if we go up to two loops, we restrict the calculations to the exchange of at most one graviton, which already involves a lot of Feynman graphs. This limitation also corresponds to restraining to study processes up to the order of $E^2$ in the energy expansion of the EFT and to consider higher-order operators (HOs) up to the order $6$, in the Lagrangian. Two-graviton exchange would mean to calculate a lot more graphs and the introduction of several HOs of the order $8$ or more. Besides that, when considering two-graviton exchange, we will end up with contributions proportional to $\kappa^4$, which are much smaller than the one-graviton exchange approximation; therefore, we neglect it in this work. We will postpone that to future studies. Expanding $g_{\mu\nu}$ around the flat metric as $g_{\mu\nu} = \eta_{\mu\nu} + \kappa h_{\mu\nu}$, with $\eta_{\mu\nu}=(+,-,-,-)$, we have \cite{Choi:1994ax} 
\begin{eqnarray}\label{L}
 \mathcal{L} = \mathcal{L}_h + \mathcal{L}_f + \mathcal{L}_A +\mathcal{L}_{HO} + \mathcal{L}_{GF} + \mathcal{L}_{CT},
\end{eqnarray}
\noindent where
\begin{subequations}\label{Lexp}
 \begin{eqnarray}
  &&\mathcal{L}_h = -\frac{1}{4}\partial_\mu h\partial^\mu h +\frac{1}{2}\partial_{\mu}h^{\sigma\nu}\partial^\mu h_{\sigma\nu};\label{Lh}\\
  &&\mathcal{L}_f = \frac{i}{2}\left(\bar{\psi}\gamma^\mu\partial_\mu\psi - \partial_\mu\bar{\psi}\gamma^\mu\psi\right) - m\bar{\psi}\psi + e\bar{\psi}\gamma^\mu\psi A_\mu\nonumber\\
  &&\quad\quad~+\frac{\kappa}{2}h\left[\frac{i}{2}(\bar{\psi}\gamma^\mu\partial_\mu\psi - \partial_\mu\bar{\psi}\gamma^\mu\psi) - m\bar{\psi}\psi\right] - \frac{\kappa}{4}h_{\mu\nu}\left(\bar{\psi}\gamma^\mu\partial\nu\psi - \partial^\nu\bar{\psi}\gamma^\mu\psi\right)\nonumber\\
  &&\quad\quad~{-\frac{1}{2}\kappa e\left(h\eta_{\mu\nu}-h_{\mu\nu}\right)\bar{\psi}\gamma^\mu\psi A^\nu};\label{Lf}\\
  &&\mathcal{L}_A = -\frac{1}{4}F^{\mu\nu}F_{\mu\nu} + \frac{\kappa}{2}h^\tau_{~\nu}F^{\mu\nu}F_{\mu\tau} - \frac{\kappa}{8}hF^{\mu\nu}F_{\mu\nu};\label{LA}\\
  &&\mathcal{L}_{HO}=i\bar\psi~\frac{\Box}{M_P^2}\left(\tilde{e}_1\slashed{\partial}- \tilde{e}_2 m\right)\psi-\frac{\tilde{e}_3}{4}F^{\mu\nu}\frac{\Box}{M_P^2} F_{\mu\nu}+ \frac{i\tilde{e}_4}{M_P^2}\bar\psi\gamma_\mu\partial_\nu\psi F^{\mu\nu};\label{LHO}\\
  &&\mathcal{L}_{GF}=\frac{1}{\xi_h}\left(\partial^\nu h_{\mu\nu}-\frac{1}{2}\partial_\mu h\right)^2 - \frac{1}{2\xi_A}(\partial_\mu A^\mu)^2\label{LGF},
 \end{eqnarray}
\end{subequations}
\noindent where $h=h^\mu_{~\mu}, F^{\mu\nu}$ is the usual electromagnetic field strength, {$\sigma^{\mu\nu}=\frac{i}{2} [\gamma^\mu,\gamma^\nu]$}, $\xi_h$ is the gravitational gauge-fixing parameter, $\xi_A$ is the electromagnetic gauge-fixing parameter and $\tilde{e}_i$ are dimensionless coupling constants related to the higher (sixth) order operators, needed at our approximations. $\mathcal{L}_{CT}$ is the Lagrangian of counterterms. They are monomials of the same exact form as that of Eq. (\ref{Lexp}) multiplied by convenient $Z$ factors, to be properly chosen to absorb the divergencies obtained in the calculations. We did not write the Faddeev-Popov Lagrangian (nor the corresponding ghost propagators, below). The electromagnetic ghosts completely decouple, as known, for Abelian gauge theories, and the gravitational ghosts are not needed in the order that we are working (only one-graviton exchange).

From the quadratic part of the Lagrangian \eqref{L}, we find the following propagators (in an arbitrary gauge):
\begin{subequations} \label{eq04}
\begin{align}
\langle TA^\mu(p) A^\nu(-p)\rangle & = \frac{i}{p^2}\left(\eta^{\mu\nu}-(1-\xi_A)\frac{p^\mu p^\nu}{p^2} \right);\\
\langle T\psi(p)\bar\psi(-p)\rangle & =i\frac{\slashed{p}-m}{p^2-m^2};\\
\langle Th^{\alpha\beta}(p) h^{\mu\nu}(-p)\rangle & =\frac{i}{p^2}\left(P^{\alpha\beta\mu\nu}-(1-\xi_h)\frac{Q^{\alpha\beta\mu\nu}}{p^2}\right),
\end{align}
\end{subequations}
where the projectors $P^{\alpha\beta\mu\nu}$ and $Q^{\alpha\beta\mu\nu} $ are given by
\begin{eqnarray}
P^{\alpha\beta\mu\nu} &=&\frac{1}{2} \left(\eta^{\alpha\mu}\eta^{\beta\nu}+\eta^{\alpha\nu}\eta^{\beta\mu}-\eta^{\alpha\beta}\eta^{\mu\nu} \right);\nonumber\\
Q^{\alpha\beta\mu\nu} &=& (\eta^{\alpha\mu}p^\beta p^\nu+\eta^{\alpha\nu}p^\beta p^\mu+\eta^{\beta\mu}p^\alpha p^\nu+\eta^{\beta\nu}p^\alpha p^\mu).
\end{eqnarray}

In the renormalized Lagrangian, we followed the usual notation for the fields renormalizations: $\psi_{0}=Z_{2}^{1/2}\psi$ and $A_{0\mu}=Z_3^{1/2}A_{\mu}$, with the renormalizing factors expanded in loops, as
\begin{eqnarray}
Z_{i}=1+ Z_{i}^{(1)}+Z_{i}^{(2)}+\cdots.
\end{eqnarray}
\noindent  The relation between the bare and renormalized electric charges is given by
\begin{eqnarray}\label{eq_e_0}
e_0&=&\mu^{2\epsilon}\frac{Z_1}{Z_2 Z_{3}^{1/2}}e,
\end{eqnarray}
\noindent where $Z_{1}^{(l)}$ is the counterterm in $l$ loops, for the interaction vertex $e\bar{\psi}\slashed{A} \psi$, while $\mu$ is a mass scale introduced by the dimensional regularization with $D=4-2\epsilon$.

\subsection{One-loop renormalization group functions}\label{sec1-oneloop}

\begin{figure}[ht]
	\includegraphics[angle=0,width=6cm]{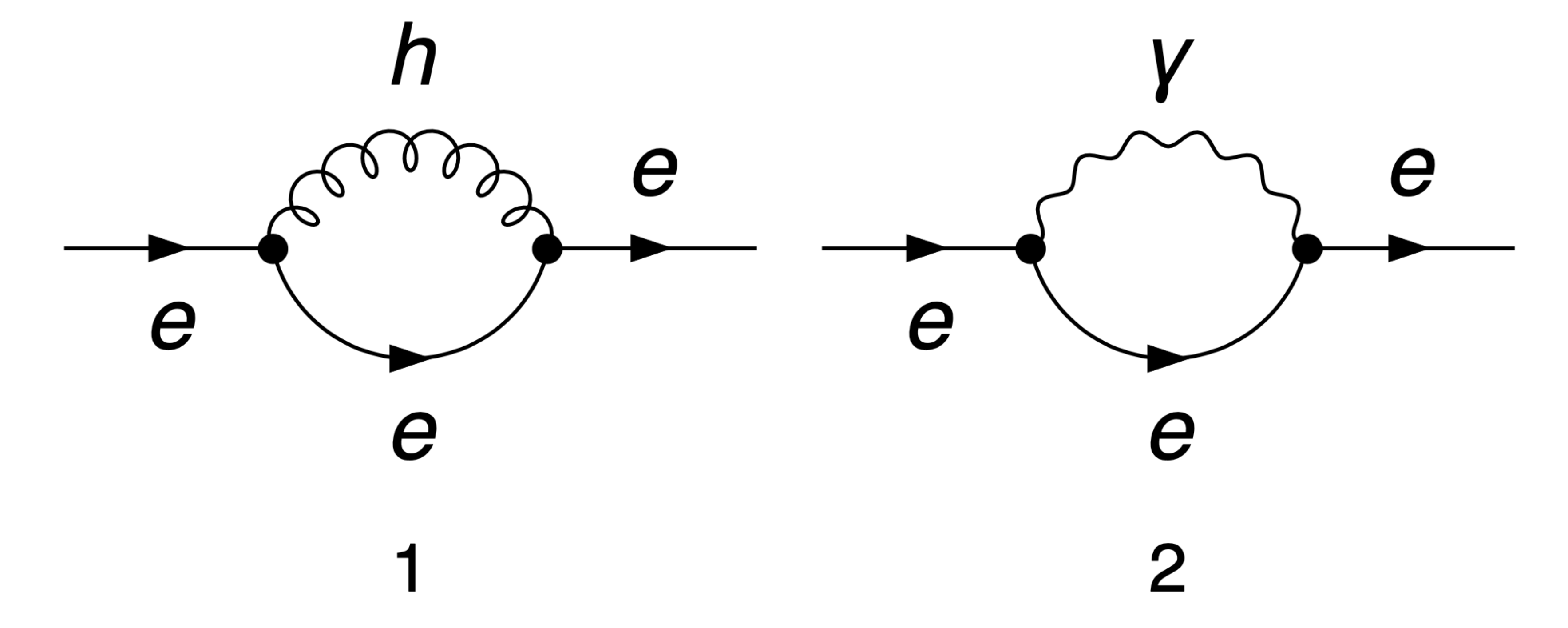}
	\caption{Feynman diagrams for the electron self-energy. Continuous, wavy and wiggly lines represent the electron, photon, and graviton propagators, respectively.}
	\label{fig01}
\end{figure}

\begin{figure}[ht]
	\includegraphics[angle=0 ,width=6cm]{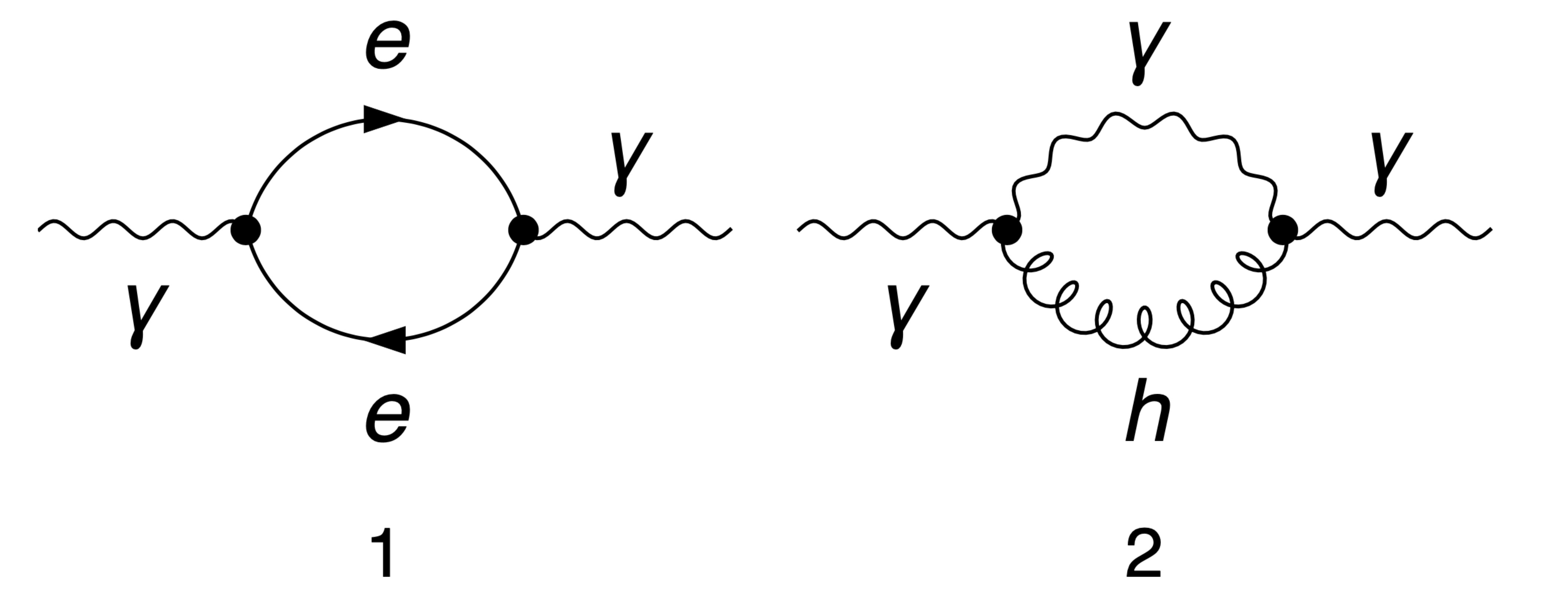}
	\caption{Feynman diagrams for the photon self-energy.}
	\label{fig02}
\end{figure}

In this section we will calculate the one-loop relevant $Z$ factors for the renormalization of the electric charge in an arbitrary gauge, so we will consider the one-loop self-energy diagrams in Figs. \ref{fig01} and \ref{fig02}. To do this we use a set of \textit{Mathematica} packages~\cite{feyncalc,feynarts,feynrules}, whose files can be found in Supplemetal Material~\cite{site_lehum}. 

The corresponding expression for the electron self-energy (Fig. \ref{fig01}) is
\begin{eqnarray}\label{1eq02}
-i\Sigma^{(1)}(p) &=& \frac{\left(32 \xi_A e^2+37\kappa^2 m^2-29\xi_h\kappa^2 m^2\right) \slashed{p}-2m \left(16 e^2 (\xi_A+3)-\kappa^2 m^2(19\xi_h-23)\right)}{512 \pi^2 \epsilon } \nonumber\\
&& + \frac{p^2\kappa^2 \slashed{p}(15\xi_h-19)+12 m(3-\xi_h)}{128 \pi^2 \epsilon } \nonumber\\
&&+ \; Z_{2}^{(1)}~\slashed{p}-m Z_{m}^{(1)} + \frac{p^2}{M_P^2}\Big(Z_{\tilde{e}_1}^{(1)}~\slashed{p}-m Z_{\tilde{e}_2}^{(1)}\Big) + \mathrm{finite}.
\end{eqnarray}

The first term of the above equation is renormalized by the counterterms $Z_2^{(1)} $ and $Z_m^{(1)} $, while the second term is renormalized by the higher derivative operators $Z_{\tilde{e}_1}^{(1)}$ and $Z_{\tilde{e}_2}^{(1)}$ (they are not needed for our conclusions and we will not write them, below).
Imposing finiteness to $\Sigma(p)$, we find the following one-loop counterterms:
\begin{subequations}\label{ct01}
\begin{eqnarray}
Z_{2}^{(1)} &=& -\frac{e^2 \xi_A}{16 \pi ^2 \epsilon }-\frac{(37-29\xi_h) \kappa ^2 m^2}{512 \pi ^2 \epsilon },\label{ct01-1}\\
Z_{m}^{(1)} &=& -\frac{3 e^2 m}{16 \pi ^2 \epsilon }-\frac{e^2 m \xi_A}{16 \pi ^2 \epsilon }-\frac{23 \kappa ^2 m^3}{256 \pi ^2 \epsilon }+\frac{19 \kappa ^2 m^3 \xi_h }{256 \pi ^2 \epsilon },\label{ct01-2}
\end{eqnarray}
\end{subequations}
\noindent where we used the minimal subtraction scheme (MS) ~\cite{minsub}.

For the photon self-energy, straight calculation of the graphs in Fig. \ref{fig02} results in
\begin{eqnarray}
\Pi^{(1)}_{\mu\nu}(p)=\left(p^2 \eta_{\mu  \nu }-p_{\mu } p_{\nu }\right)\Pi^{(1)}(p^2),
\end{eqnarray}
\noindent where
\begin{eqnarray}
\label{eq_pi1}
\Pi^{(1)}(p^2)= Z_3^{(1)}+\tilde{Z}_3^{(1)}\frac{p^2}{M_P^2}+\frac{\left(8 e^2-\kappa^2 p^2(2-3\xi_h)\right)}{96 \pi ^2 \epsilon }+\mathrm{finite}.
\end{eqnarray}

As can be seen, $\Pi^{(1)}_{\mu\nu}(p)$ preserves the Lorentz invariance and the gauge symmetry, that is expressed by the Ward identity $p^{\mu} \Pi^{(1)}_{\mu\nu}(p)=0$.
From the expression for $\Pi^{(1)}(p^2)$, we can see that $Z_3$ is the renormalizing factor for the Maxwell term, while $\tilde{Z}_3$ renormalizes the higher derivative term $F^{\mu\nu}\Box F_{\mu\nu}$. Thus, $Z_3$ is the relevant counterterm for the beta function of the electric charge. 

Imposing finiteness of $\Pi^{(1)}(p^2)$ order by order in powers of $p^2$, we find (in the MS scheme of renormalization) 
\begin{eqnarray}\label{oneloopZ3}
Z_3^{(1)} &=& -\frac{e^2}{12\pi^2\epsilon},\\
\tilde{Z}_3^{(1)} &=& \frac{\kappa^2M_P^2(2-3\xi_h)}{96\pi^2\epsilon}.
\end{eqnarray}
\noindent 

\begin{figure}[ht]
	\includegraphics[angle=0 ,width=12cm]{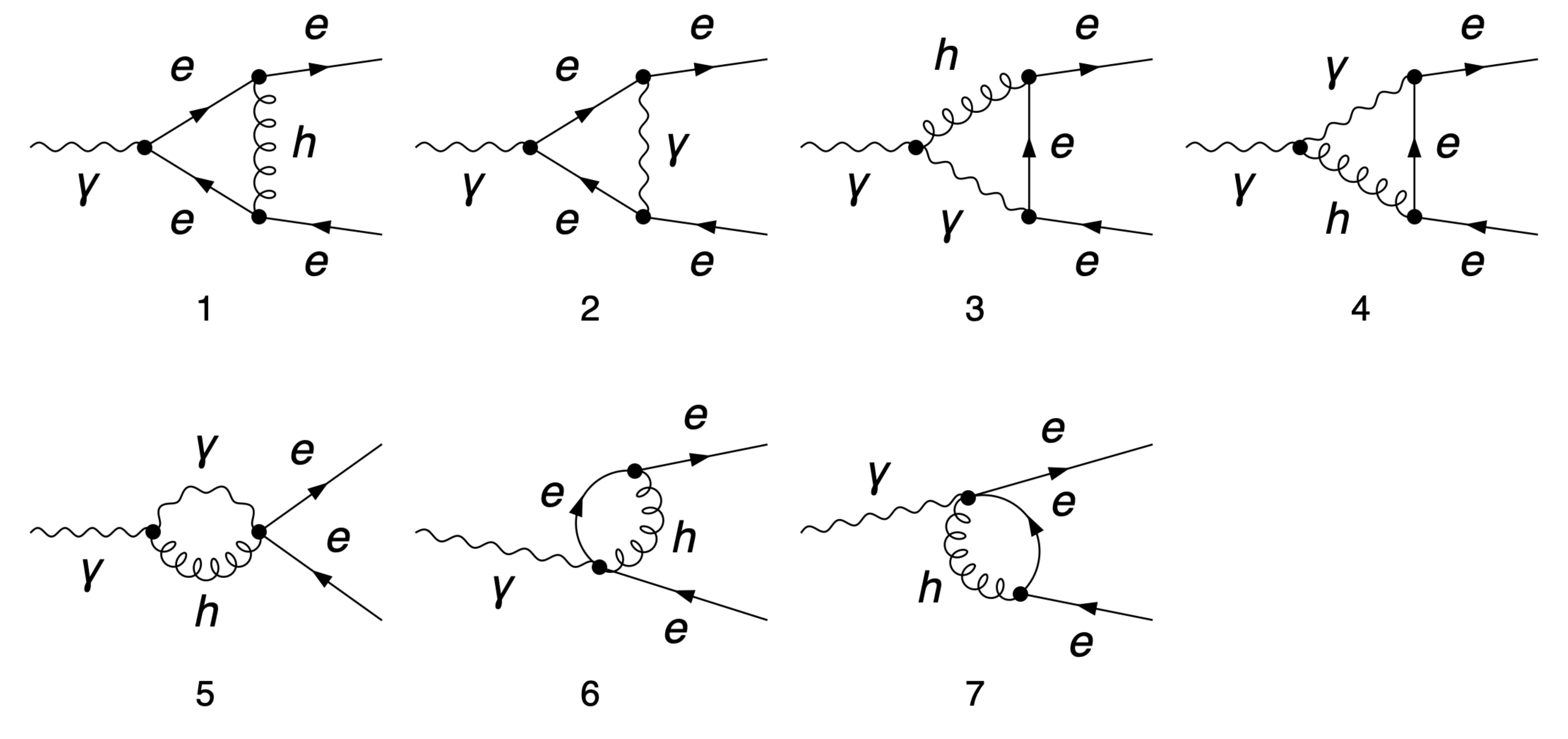}
	\caption{One-loop corrections to the $\bar\psi\gamma^\mu\psi A_{\mu}$ vertex function.}
	\label{vertex_qed}
\end{figure}

Note that $Z_3^{(1)}$ is independent of the gravitational coupling $\kappa$.
Let us go to the vertex function $\Gamma^{\mu}$. The contributions to it, in one-loop order, are depicted in Fig. \ref{vertex_qed}. The resulting expression is
\begin{eqnarray}\label{vertex1}
-i\Gamma^{(1)}_{\mu}(p) &=& -e\gamma_\mu\left[Z_{1}^{(1)}+\frac{e^2\xi_A}{16 \pi^2 \epsilon}+\frac{\kappa^2 m^2(37-29\xi_h)}{512 \pi^2 \epsilon }\right]\nonumber\\
&&-2i\tilde{e}\,\tilde{Z}_{1}^{(1)}\,\sigma_{\mu\nu} p^{\nu}+\mathcal{O}(p) +\mathrm{finite}.
\end{eqnarray}

$\tilde{Z}_1^{(1)}$ absorbs the terms linear in $p^{\mu}$ (not shown in the above expression). Imposing finiteness through the MS scheme, we find
\begin{eqnarray}\label{ct031}
Z_{1}^{(1)} &=&-\frac{e^2 \xi_A}{16 \pi ^2 \epsilon }-\frac{(37-29\xi_h) \kappa ^2 m^2}{512 \pi ^2 \epsilon }. 
\end{eqnarray}

Note that $Z_{1}^{(1)}=Z_{2}^{(1)}$ see Eq.\eqref{ct01-1}, confirming the Ward identity: $p^{\mu} \Gamma_{\mu}(p)= \Sigma(q)-\Sigma(p+q)$ (see, for instance, Ref. \cite{Srednicki:2007qs}), required by gauge symmetry. This result shows that, up to the order of one loop and $\kappa^2$, gravitational corrections do not spoil the gauge invariance.

The beta function of the electric charge can now be obtained from the relation between bare and renormalized electric charge Eq.(\ref{eq_e_0}). Observe that $Z_1^{(1)}$ and $Z_2^{(1)}$ depend on $\kappa$, but, since $Z_{1}^{(1)}=Z_{2}^{(1)}$, the renormalization of $e_0=\mu^{2\epsilon}Z_{3}^{-1/2}e$, is independent of $\kappa$ and $\beta(e)$ at the one-loop order results in
\begin{eqnarray}\label{beta_e_qed1l}
\beta^{(1)}(e)=\mu\frac{de}{d\mu}=\frac{e^3}{12\pi^2},
\end{eqnarray}
\noindent which is independent of $\kappa$, just as in our previous result for the Einstein-scalar QED in one loop ~\cite{Bevilaqua:2015hma}.
 
\subsection{Two-loop corrections
 of the electric charge} \label{sec1-twoloop}

\begin{figure}[ht]
	\includegraphics[angle=0 ,width=12cm]{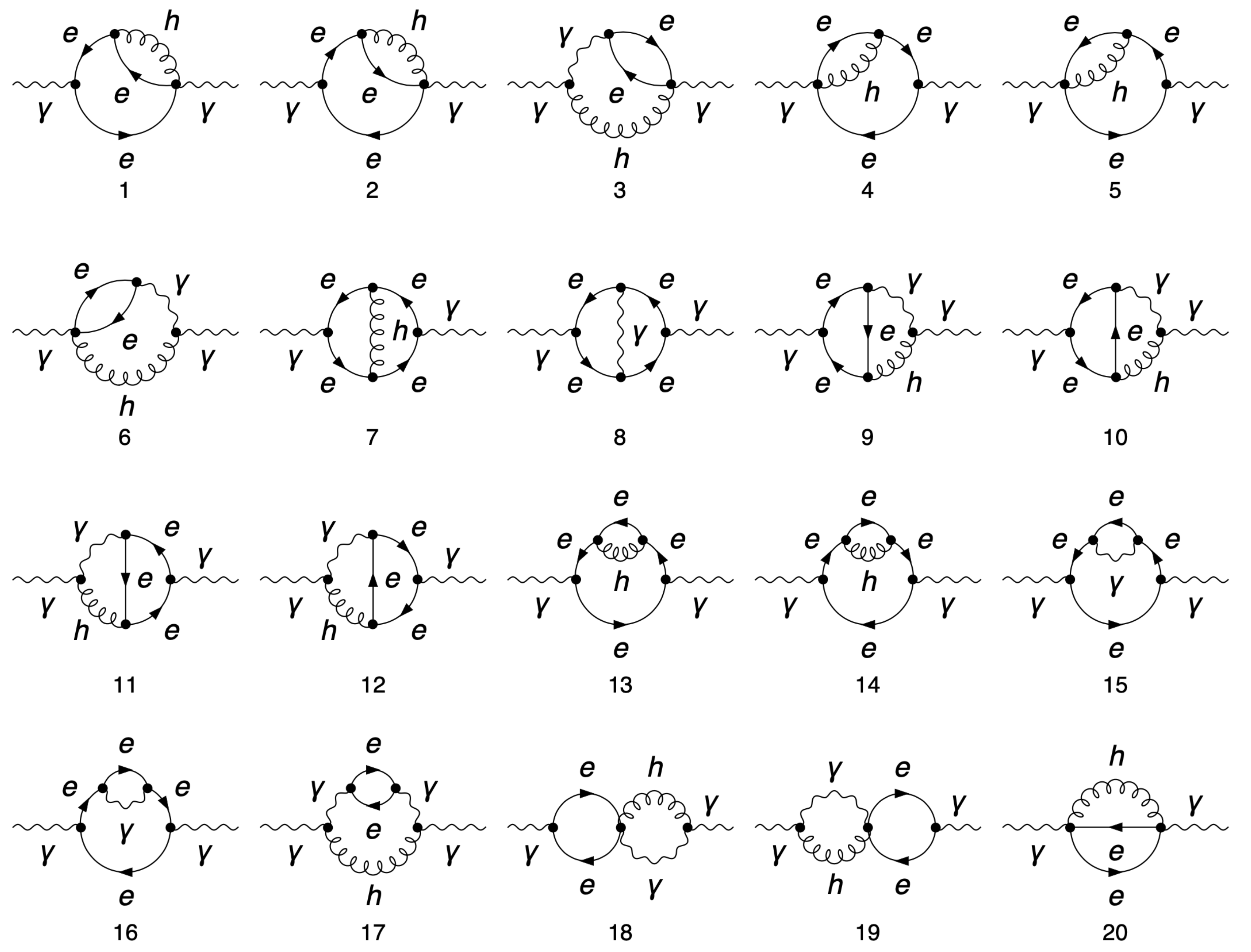}
	\caption{Two-loop corrections to the photon self-energy.}
	\label{fig03}
\end{figure}

In order to compute the two-loop corrections to the beta function of the electric charge, we have to evaluate the two-loop corrections to the photon self-energy $\Pi^{(2)}_{\mu\nu}$,  besides the two-loop scattering amplitude $\Gamma^{(2)}_{\mu}$ and the electron self-energy $\Sigma^{(2)}$. The diagrams contributing to $\Pi^{(2)}_{\mu\nu}$ are depicted in 
Fig. \ref{fig03}.
The Feynman integrals were calculated using the set of computational packages given in the Refs.~\cite{feyncalc,feynarts,feynrules,Mertig:1998vk}. 

We already used this computational package to calculate the Lorentz tensor integrals in one loop. In two loops, it  permits to calculate only Lorentz scalar integrals. So, for each Feynman graph $i$ contributing to  $\Pi^{(2)\mu\nu}_{i}(p)$, we write the general Lorentz symmetric form:
\begin{equation}
\Pi^{(2)\mu\nu}_i(p)=\eta^{\mu\nu}p^2~\Pi^{(2)}_i(p^2)+p^\mu p^\nu \tilde{\Pi}^{(2)}_i(p^2),
\end{equation} 
from  which $\Pi^{(2)}_i(p^2)$ and $\tilde{\Pi}^{(2)}_i(p^2)$ can be obtained through the projections
\begin{eqnarray}
\Pi^{(2)}_{i} &=& \frac{1}{(D-1)p^2}\left(\eta_{\mu\nu}-\frac{p_{\mu}p_\nu}{p^2}\right){\Pi_{i}^{(2)\mu\nu}},\nonumber\\ 
\tilde{\Pi}^{(2)}_i &=& -\frac{1}{(D-1)p^2}\left(\eta_{\mu\nu}-D\frac{p_{\mu}p_\nu}{p^2}\right)\Pi_i^{(2)\mu\nu}. \nonumber
\end{eqnarray}
As defined before, $D=4-2\epsilon$. After calculating $\Pi_i^{(2)}$ and $\tilde{\Pi}_i^{(2)}$ for the several diagrams and adding the results, we found that their sum satisfies $\Pi^{(2)}(p^2)=-\tilde{\Pi}^{(2)}(p^2)$. This result implies that the two-loop photon polarization tensor has the transversal form, compatible with the Ward identity $p^{\mu} \Pi^{(2)}_{\mu\nu}=0$ and the preservation of the gauge symmetry.

In those calculations, we reduced the scalar two-loop integrals using the Tarasov algorithm~\cite{Tarasov:1997kx}, with the help of the computational package TARCER~\cite{Mertig:1998vk}. The resulting scalar two-loop integrals can be found in Ref. \cite{Martin:2003qz}. Finally, we evaluated the integrals, keeping only the UV-divergent part of $\Pi^{(2)}(p^2=0)$. 

The detailed file with the calculations can be found in Supplemental Material~\cite{site_lehum}, and the resulting expressions for the UV-divergent part is given by
\begin{eqnarray}
-i\Pi^{(2)}(p)=\frac{3 e^4}{128 \pi ^4 \epsilon }
-\frac{5 e^2 \kappa ^2 m^2}{768 \pi ^4 \epsilon }
+\mathcal{O}\left(p^2\right).
\end{eqnarray}
Corroborating the gauge independence of this result, we see that this expression does not depend on the gauge parameters $\xi_A$ and $\xi_h$. This result provides the renormalization of the Maxwell term. It also contributes for the two-loop correction to the beta function of the electric charge, through the renormalization constant $Z^{(2)}_3$ see Eq. (\ref{eq_e_0}).

To get the $\beta$ function of $e$ in two loops, we would also need $Z^{(2)}_2$ and $Z^{(2)}_1$, which means to calculate the electron self-energy $\Sigma^{(2)}$ and the electron-photon scattering amplitude $\Gamma^{(2)}_{\mu}$ in two loops. In the previous section, we  calculated these amplitudes in one loop, in an arbitrary gauge, for both electromagnetism and gravitation, with the result that $Z_1^{(1)}=Z_2^{(1)}$. From now on, we will assume (without proving) that the same result $Z_1^{(2)}=Z_2^{(2)}$  remains true in two loops. This means that the relation between $e_0$ and $e$ remains given by $e_0=\mu^{2\epsilon}Z_{3}^{-1/2}e$, in two loops.

\begin{figure}[ht]
	\includegraphics[angle=0 ,width=8cm]{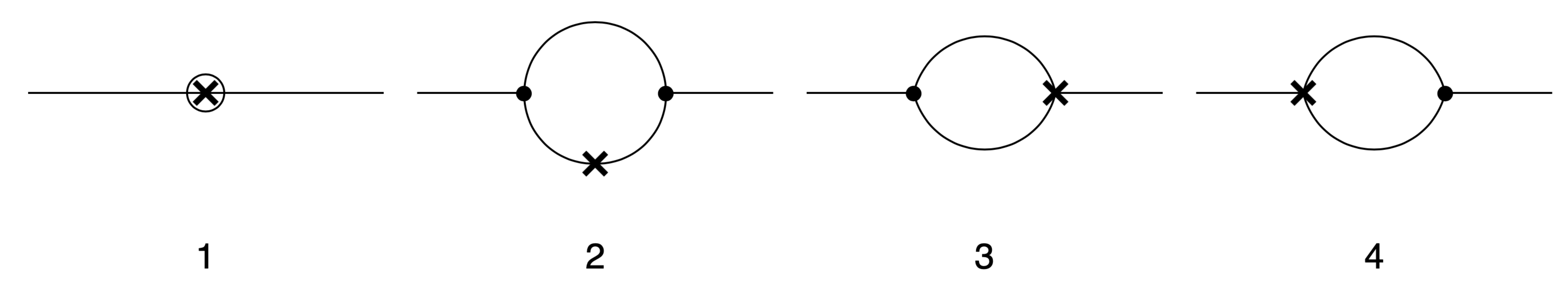}
	\caption{Topologies of the diagrams of counterterms in the Einstein-QED model.}
	\label{fig04}
\end{figure}

Let us determine $Z_3^{(2)}$. The topologies of the counterterm diagrams are shown in Fig. \ref{fig04}, and the corresponding expression up to the order of $e^4$ is given by
\begin{eqnarray}
-i\Pi_{CT} &=& -Z_3^{(2)}+\frac{e^2}{6 \pi^2} \left(Z_m^{(1)}-Z_1^{(1)}\right)\nonumber\\
&=& -Z_3^{(2)}-\frac{e^4}{32 \pi ^4 \epsilon } +\mathcal{O}\left(p^2\right),
\end{eqnarray}
\noindent where $Z_3^{(2)}$ is the two-loop order counterterm.

The divergent part of the two-loop photon self-energy has to be canceled by the counterterm $Z_3^{(2)}$, so we write
\begin{eqnarray}
-i\Pi_{UV}(p) &=&-i	\Pi_{CT}(p)-i\Pi_2(p)\nonumber\\
&=&-Z_3^{(2)} -\frac{e^4}{128 \pi ^4 \epsilon }-\frac{5 e^2 \kappa ^2 m^2}{768 \pi ^4 \epsilon }
+\mathcal{O}\left(p^2\right)=0.
\end{eqnarray}

Therefore, the renormalizing factor for the gauge field $Z_3$, at two-loop order, is given by 
\begin{eqnarray}
Z_3 &=& Z_3^{(0)} + Z_3^{(1)}+ Z_3^{(2)}+\cdots \nonumber\\
&=& 1 -\frac{e^2}{12 \pi ^2 \epsilon }-\frac{e^4}{128 \pi ^4 \epsilon }-\frac{5 e^2 \kappa ^2 m^2}{768 \pi ^4 \epsilon }+\cdots,
\end{eqnarray}
\noindent where we have used Eq. \eqref{oneloopZ3}. Notice that, in the absence of gravity ($\kappa\rightarrow0$), our result agrees with previous calculations found in the literature (see, for instance,~\cite{Honemann:2018mrb}).

The corrections up to two loops and  up to the order of $\kappa^2$ to the $\beta$ function of the electric charge can, thus, be cast as 
\begin{eqnarray}
\beta(e)&=&\mu \frac{\delta e}{\delta\mu}=\frac{e^3}{12\pi^2}+\frac{e^5}{128\pi^4}+\frac{5e^3\kappa^2m^2}{768\pi^4}
\end{eqnarray}
\noindent or, in terms of the fine-structure constant $\alpha=e^2/4\pi$,
\begin{eqnarray}
\beta(\alpha)=&&\beta(e)\frac{d \alpha}{d e}=\frac{2\alpha^2}{3 \pi}\left(1+\frac{5}{2\pi}\frac{m^2}{M_P^2}\right)+\frac{\alpha^3}{4 \pi^2},
\end{eqnarray}
\noindent where $M_P$ is the Planck mass. Since $M_P\approx 1.22\times10^{19}\text{~}\mathrm{GeV}$ and the mass of the electron is $m\approx 0.51\text{~}\mathrm{MeV}$, the gravitational correction to the beta function at $\alpha^2$ order should be of the order of $\frac{m^2}{M_P^2}\sim10^{-45}$. As we see, even though $\beta(e)$ becomes dependent on $\kappa$ through the two-loop diagrams, the gravitational corrections do not qualitatively alter the behavior of the running of the electric charge.  

In the next section, we will compute the two-loop beta function of the electric charge, in scalar QED, including gravitational corrections up to the order of $\kappa^2$ and show that it does not qualitatively alter its usual behavior.

\section{Scalar electrodynamics coupled to gravity}\label{sec2}

We now consider the model described by
\begin{eqnarray}\label{sQED}
\mathcal{L}=&& \sqrt{-g}\Big\{\frac{2}{\kappa^2}R-\frac14 g^{\mu\nu}g^{\alpha\beta} F_{\mu\alpha} F_{\nu\beta} - g^{\mu\nu}(D_{\mu}\phi)^\dagger D_{\nu}\phi - m^2\phi^\dagger\phi - \frac{\lambda}{4} (\phi^\dagger\phi)^2 \Big\} \nonumber \\
&&\qquad\qquad +\mathcal{L}_{HO} +\mathcal{L}_{GF}+\mathcal{L}_{CT},
\end{eqnarray}
\noindent where $D_{\mu}=\partial_{\mu} - ieA_{\mu}$ is the covariant derivative and, as is well known, we have the additional $\lambda (\phi^\dagger \phi)^2$ self-interaction of the scalar field, required by renormalization.

As done in the last section, we will expand the metric around the flat metric as $g_{\mu\nu}=\eta_{\mu\nu}+\kappa h_{\mu\nu}$. By doing so, we obtain \cite{Choi:1994ax}
\begin{equation}
 \mathcal{L} = \mathcal{L}_h + \mathcal{L}_s + \mathcal{L}_A + \mathcal{L}_{HO} + \mathcal{L}_{GF} + \mathcal{L}_{CT},
\end{equation}
\noindent where the parts $\mathcal{L}_h,\mathcal{L}_A$, and $\mathcal{L}_{GF}$ are given by Eqs. \eqref{Lh},\eqref{LA}, and \eqref{LGF} respectively, and
\begin{subequations}
 \begin{eqnarray}
  \mathcal{L}_s &=& (D^\mu\phi)^\dagger D_\mu\phi - m^2(\phi^\dagger\phi) -\frac{\lambda}{4}(\phi^\dagger\phi)^2 - \kappa h^{\mu\nu}(D_\mu\phi)^\dagger D_\nu\phi\nonumber\\
  &&+ \frac{\kappa}{2}h\left[(D^\mu\phi)^\dagger D_\mu\phi - m^2 \phi^\dagger\phi -\frac{\lambda}{4}(\phi^\dagger\phi)^2\right];\\
  \mathcal{L}_{HO} &=& \frac{\tilde{\lambda}_1}{M_P^2}\left[\mathrm{Re}(\phi^*\partial_\mu\phi)\right]^2
  +\frac{\tilde{\lambda}_2}{M_P^2}\left[\mathrm{Im}(\phi^*\partial_\mu\phi)\right]^2-\frac{\tilde{e}_3}{4}F^{\mu\nu}\frac{\Box}{M_P^2} F_{\mu\nu},
 \end{eqnarray}
\end{subequations}
\noindent where we have used the same conventions as the last section.

The propagators of the photon and the graviton are given by Eq. (\ref{eq04}), and the scalar propagator is given by
\begin{eqnarray}\label{eq-prop-phi}
\langle T\phi(p)\phi^\dagger(-p)\rangle &=&\frac{i}{p^2-m^2}.
\end{eqnarray}

The relations between bare and renormalized coupling constants are given by
\begin{eqnarray}\label{eq_SQED_e0}
e_0&=&\mu^{2\epsilon}\frac{Z_1}{Z_2 Z_{3}^{1/2}}e,\label{eq_SQED_e00} \\
\lambda_0&=&\mu^{2\epsilon}\frac{Z_\lambda}{Z_2^2}\lambda \label{eq_SQED_l0},
\end{eqnarray}
\noindent where $Z_1$ and $Z_2$ are, respectively, the renormalization factors for the interaction term $ie A^{\mu}(\phi^\dagger\partial_\mu\phi-\phi\partial_\mu\phi^\dagger)$ and the scalar field while $Z_\lambda$ is the counterterm for the scalar self-interaction.

We now proceed to discuss the one- and two-loop corrections to the scalar and photon self-energies in order to compute the relevant $Z$ factors.

\subsection{One-loop renormalization group functions} \label{sec2-oneloop}

\begin{figure}[ht]
	\includegraphics[angle=0 ,width=10cm]{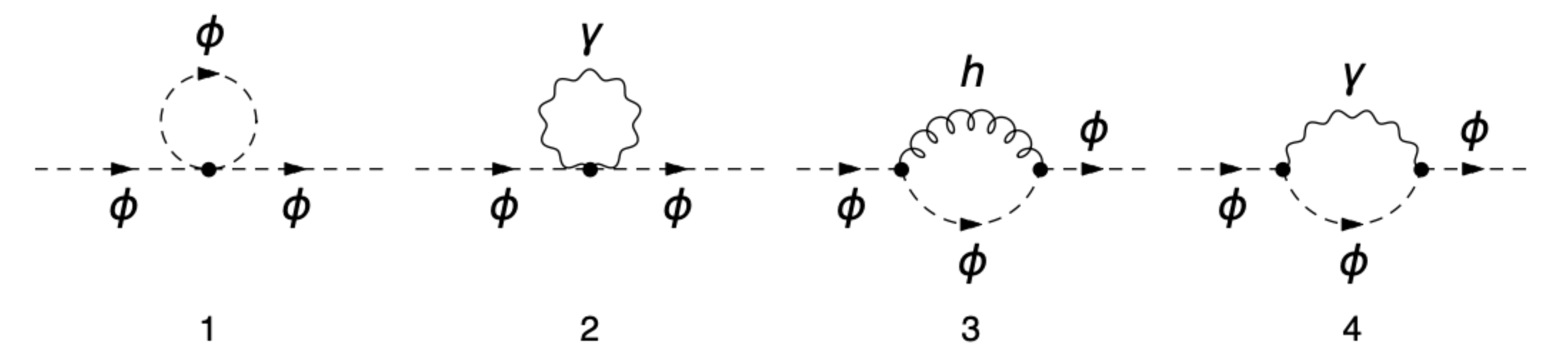}
	\caption{Self-energy of the scalar field. Dashed, wavy, and wiggly lines represent the scalar field, photon and graviton propagators, respectively.}
	\label{fig05}
\end{figure}

Let us start with the one-loop scalar self-energy (Fig. \ref{fig05}). The corresponding expression is given by
\begin{eqnarray}
 \Sigma(p) &=& \frac{m^2 \left(\lambda-e^2\xi _{A}-\kappa^2 m^2(2- \xi _{h})\right)}{16 \pi ^2 \epsilon }-\frac{p^2 \left(e^2 (3-\xi _{A})-\kappa ^2 m^2 (2- \xi_{h})\right)}{16 \pi ^2 \epsilon }\nonumber\\
&& + Z_2^{(1)}p^2 - Z_m^{(1)}m^2+\mathrm{HO~terms}.
\end{eqnarray}
\noindent Therefore, imposing finiteness through the MS we have
\begin{subequations}
 \begin{eqnarray}
  &&Z_2^{(1)} = \frac{e^2 (3-\xi_{A})-\kappa ^2 m^2(2- \xi_{h})}{16 \pi ^2 \epsilon };\label{31a}\\
  &&Z_m^{(1)} = \frac{\lambda-e^2\xi _{A}-\kappa ^2 m^2 (2-\xi _{h})}{16 \pi ^2 \epsilon }.
 \end{eqnarray}
\end{subequations}

\begin{figure}[ht]
	\includegraphics[angle=0 ,width=8cm]{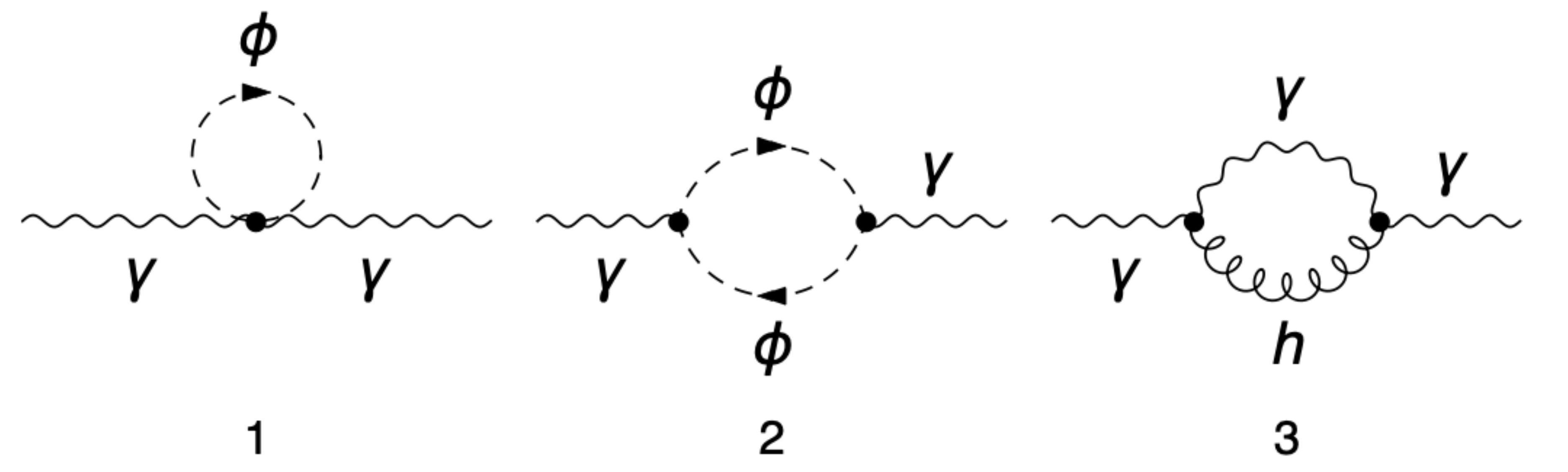}
	\caption{The one-loop photon polarization tensor in the scalar QED coupled to gravity.}
	\label{fig06}
\end{figure}

For the one-loop photon self-energy (Fig. \ref{fig06}), we found the expression \cite{Bevilaqua:2015hma}
\begin{eqnarray}
-i\Pi^{\mu\nu}(p) &=& -\left(p^2 \eta^{\mu  \nu }-p^{\mu } p^{\nu }\right)\left[Z_3^{(1)}+ \tilde{Z}_3^{(1)}\frac{p^2}{M_P^2} +\frac{e^2}{48 \pi ^2 \epsilon} +\frac{\kappa ^2 p^2 \left(2-3 \xi _{h}\right)}{96 \pi ^2 \epsilon } +\mathrm{finite}\right].
\end{eqnarray}
\noindent Imposing finiteness, the counterterms can be cast as
\begin{eqnarray}\label{1-loop_Z3_sqed}
Z_3^{(1)} &=& -\frac{e^2}{48 \pi ^2 \epsilon },\\
\tilde{Z}_3^{(1)} &=&- \frac{\kappa ^2 M_P^2 \left(2-3 \xi _{h}\right)}{96 \pi ^2 \epsilon },
\end{eqnarray}
\noindent where $Z_3^{(1)}$ is the Maxwell counterterm and $\tilde{Z}_3^{(1)}$ is a counterterm that renormalizes the high-order derivative term $F^{\mu\nu}\Box F_{\mu\nu}$.

\begin{figure}[ht]
	\includegraphics[angle=0 ,width=12cm]{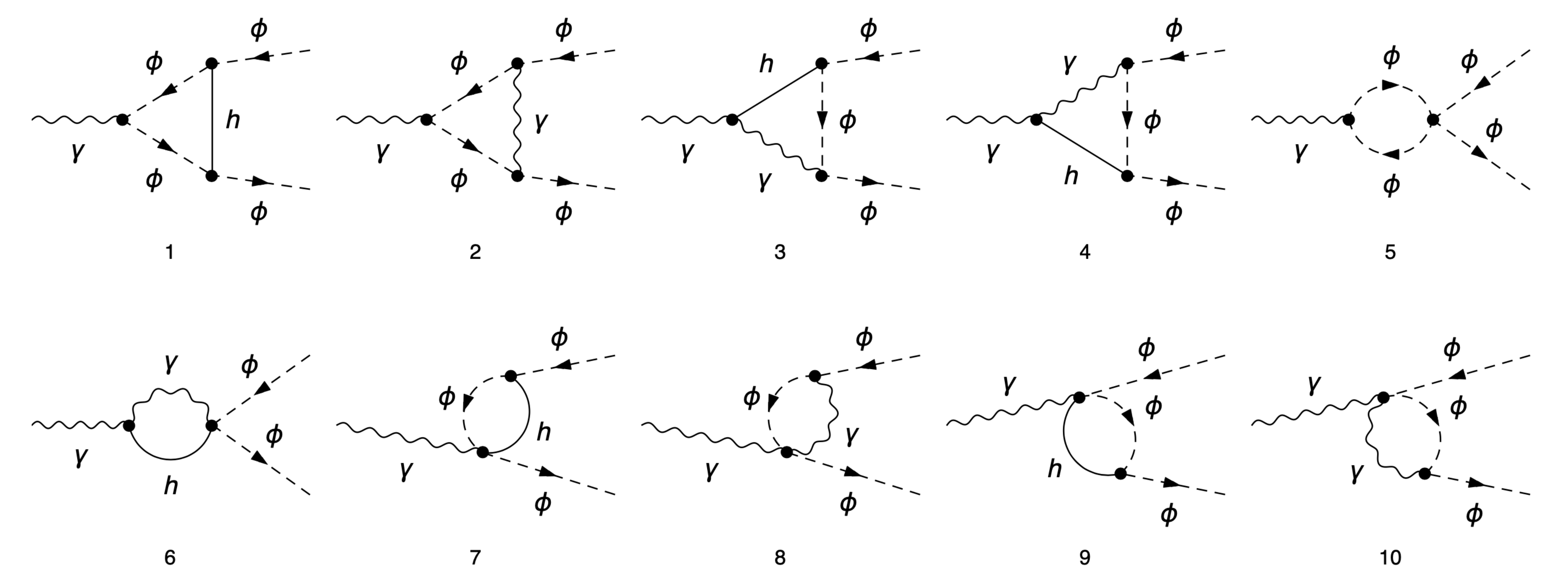}
	\caption{One-loop corrections to the $j^\mu A_{\mu}$ vertex function in the scalar QED.}
	\label{vertex_sqed}
\end{figure}

Finally, we compute the three-point function diagrams depicted in Fig. \ref{vertex_sqed}. We obtain the following expression:
\begin{eqnarray}
 -i\Gamma^\mu &=& e \left(p_1^{\mu }-2 p_2^{\mu }\right)\left[Z_1^{(1)} +  \frac{e^2 \left(\xi _{A}-3\right)-\kappa ^2 m^2 \xi _{h}+2 \kappa ^2 m^2}{16 \pi ^2 \epsilon }\right]+\mathrm{finite}.
\end{eqnarray}
\noindent Then, imposing finiteness, we find
\begin{equation}
 Z_1^{(1)} = \frac{e^2 (3-\xi_{A})-\kappa ^2 m^2(2- \xi_{h})}{16 \pi ^2 \epsilon }.\label{eq36}
\end{equation}
\noindent From the equation above, we can see that $Z_1^{(1)}=Z_2^{(1)}$, verifying the Ward identity.
Substituting this result in Eq. \eqref{eq_SQED_e00}, we get $e_0=\mu^{2\epsilon}Z_{3}^{-1/2}e$, from which the $\beta(e)$ at one-loop order results in
\begin{eqnarray}\label{beta_e_sqed1l}
\beta(e)=\mu\frac{de}{d\mu}=\frac{e^3}{48\pi^2},
\end{eqnarray}
\noindent which is independent of $\kappa$~\cite{Bevilaqua:2015hma}. It is worth to mention that we also have no gravitational contribution of the order of $\kappa^2$ for the renormalization of Maxwell's term at the one-loop order \cite{Bevilaqua:2015hma}.

\subsection{Two-loop beta function of the electric charge} \label{sec2-twoloop}

In order to compute the gravitational corrections to the two-loop beta function of the electric charge in the Einstein-scalar QED, we have to evaluate the two-loop photon self-energy. Here again, all the results are in arbitrary electromagnetic and gravitational gauges. The topologies of the graphs contributing to the process are depicted in Fig. \ref{fig08}. From these topologies, we construct 40 diagrams. The detailed calculations can be found in Supplemental Material \cite{site_lehum}. The resulting expression for the UV-divergent part is given by
\begin{eqnarray}
-i\Pi_2(p)= -\frac{e^4}{256 \pi ^4 \epsilon }-\frac{e^2 \lambda }{384 \pi ^4 \epsilon }-\frac{e^2 \kappa ^2 m^2}{256 \pi ^4 \epsilon }
+\mathcal{O}\left(p^2\right)	.
\end{eqnarray}

In Fig. \ref{fig04b}, we show the topologies of counterterm diagrams. The resulting  expression up to the order of $e^4$ is
\begin{eqnarray}
-i\Pi_{CT} &=& -Z_3^{(2)}+\frac{e^2}{48 \pi^2} \left(Z_m^{(1)}-Z_1^{(1)}\right)\nonumber\\
&=& -Z_3^{(2)}+\frac{2 e^2 \lambda -3 e^4}{768 \pi ^4 \epsilon } +\mathcal{O}\left(p^2\right),
\end{eqnarray}
\noindent where $Z_3^{(2)}$ is the two-loop order counterterm.

The vanishing of the divergent part of the two-loop photon self-energy,
\begin{eqnarray}
-i\Pi_{UV}(p) &=&	-i\Pi_{CT}(p)-i\Pi_2(p)\nonumber\\
&=&-Z_3^{(2)} -\frac{e^4}{128 \pi ^4 \epsilon }-\frac{e^2 \kappa ^2 m^2}{256 \pi ^4 \epsilon }
+\mathcal{O}\left(p^2\right)=0,
\end{eqnarray}
\noindent together with Eq.\eqref{1-loop_Z3_sqed}, gives us
\begin{eqnarray}
Z_3 &=& 1+ Z_3^{(1)}+ Z_3^{(2)}+\cdots \nonumber\\
&=& 1 -\frac{e^2}{48 \pi ^2 \epsilon }-\frac{e^4}{128 \pi ^4 \epsilon }-\frac{e^2 \kappa ^2 m^2}{256 \pi ^4 \epsilon }+\cdots.
\end{eqnarray}

In Eqs. (\ref{31a}) and (\ref{eq36}) we explicitly verified that, as in the spinorial QED,  $Z_1^{(1)}=Z_2^{(1)}$. From Eq. (\ref{eq_SQED_e00}), assuming, as we did in the spinorial QED, that $Z_1^{(2)}=Z_2^{(2)}$, the gravitational correction to the two-loop beta function of the electric charge up to the order of $\kappa^2$ can be calculated. The result is
\begin{eqnarray}
\beta(e)=&&\frac{e^3}{48 \pi ^2}+\frac{e^5}{128 \pi^4 }+\frac{e^3 \kappa^2 m^2}{256 \pi ^4}
\end{eqnarray}
\noindent or, in terms of fine-structure constant,
\begin{eqnarray}
\beta(\alpha)=&&\frac{\alpha^2}{6 \pi}\left(1+\frac{6}{\pi}\frac{m^2}{M_P^2}\right)+\frac{\alpha^3}{4 \pi^2}.
\end{eqnarray}
\noindent It is easy to see that, just as in the fermionic QED, although there is a nonzero gravitational contribution, it does not change the sign of the beta function and, therefore, does not qualitatively alter the behavior of the running of the electric charge in the scalar QED.

\section{Final Remarks}\label{summary}
 
In summary, we have computed the gravitational corrections to the two-loop beta function of the electric charge in QED and scalar QED, up to two loops, restricted to the order of $\kappa^2$, i.e., considering processes involving at most one-graviton exchange. We have shown that, even though the beta function of the electric charge receives a gravitational correction at two-loop order, this gravitational correction does not alter its qualitative behavior.

Functional renormalization group methods have been used to study the gravitational interaction in the presence of Abelian fields~\cite{Harst:2011zx,Eichhorn:2017lry,deBrito:2020dta}. In Ref.\cite{deBrito:2020dta}, the authors of the article used this approach to obtain the beta function of the gauge coupling as
\begin{eqnarray}
\beta(e)=\frac{e^3N_F}{24\pi^2}+\left(-\frac{5\hat{G}}{9\pi(1-2\hat{\Lambda})}+\frac{5\hat{G}}{18\pi(1-2\hat{\Lambda})^2}\right)e,
\end{eqnarray}
\noindent where $N_F$ is the number of fermions and $\hat{G}$ and $\hat{\Lambda}$ are the dimensionless Newton's gravitational constant and the dimensionless cosmological constant, respectively. These gravitational dimensionless parameters are defined from the ordinary ones, $G$ and $\Lambda$, as $\hat{G}=k^2G$ and $\hat\Lambda=k^4\Lambda$, where $k$ is the functional renormalization group  momentum scale. This rescaling of Newton's constant seems to be responsible for gravitational correction showing up in a lower order in $\alpha$. A deeper study is needed to elucidate the relationship between these two approaches. This will be a task for an upcoming work.   

Previous works have indicated that the presence of another dimensionful parameter, the cosmological constant \cite{Toms:2008dq}, through the dimensionless combination $\kappa^2\Lambda$, might render a negative new term to the beta function of the electric charge. Our tools can be modified to include the cosmological constant in the calculations, and we plan to pursue it in future investigation.

\acknowledgments
The authors thank M. Gomes for the useful comments. A. J. S. is partially supported by Conselho Nacional de Desenvolvimento Cient\'{\i}fico e Tecnol\'{o}gico (CNPq) under Project No. 306926/2017-2. H.S. is partially supported by Coordena\c{c}\~ao de Aperfei\c{c}oamento de Pessoal de N\'ivel Superior (CAPES).

\newpage

\begin{figure}[ht]
	\includegraphics[angle=0 ,width=12cm]{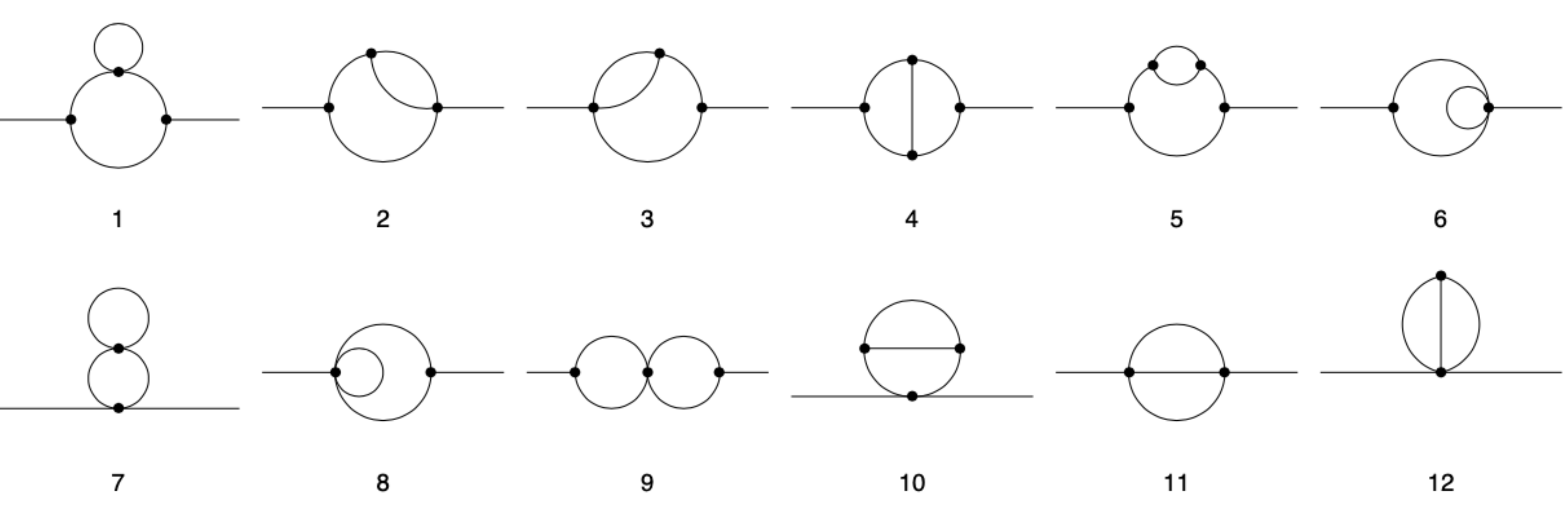}
	\caption{Topologies of the two-loop corrections to the photon self-energy in Einstein-scalar QED.}
	\label{fig08}
\end{figure}

\begin{figure}[h!]
	\includegraphics[angle=0 ,width=8cm]{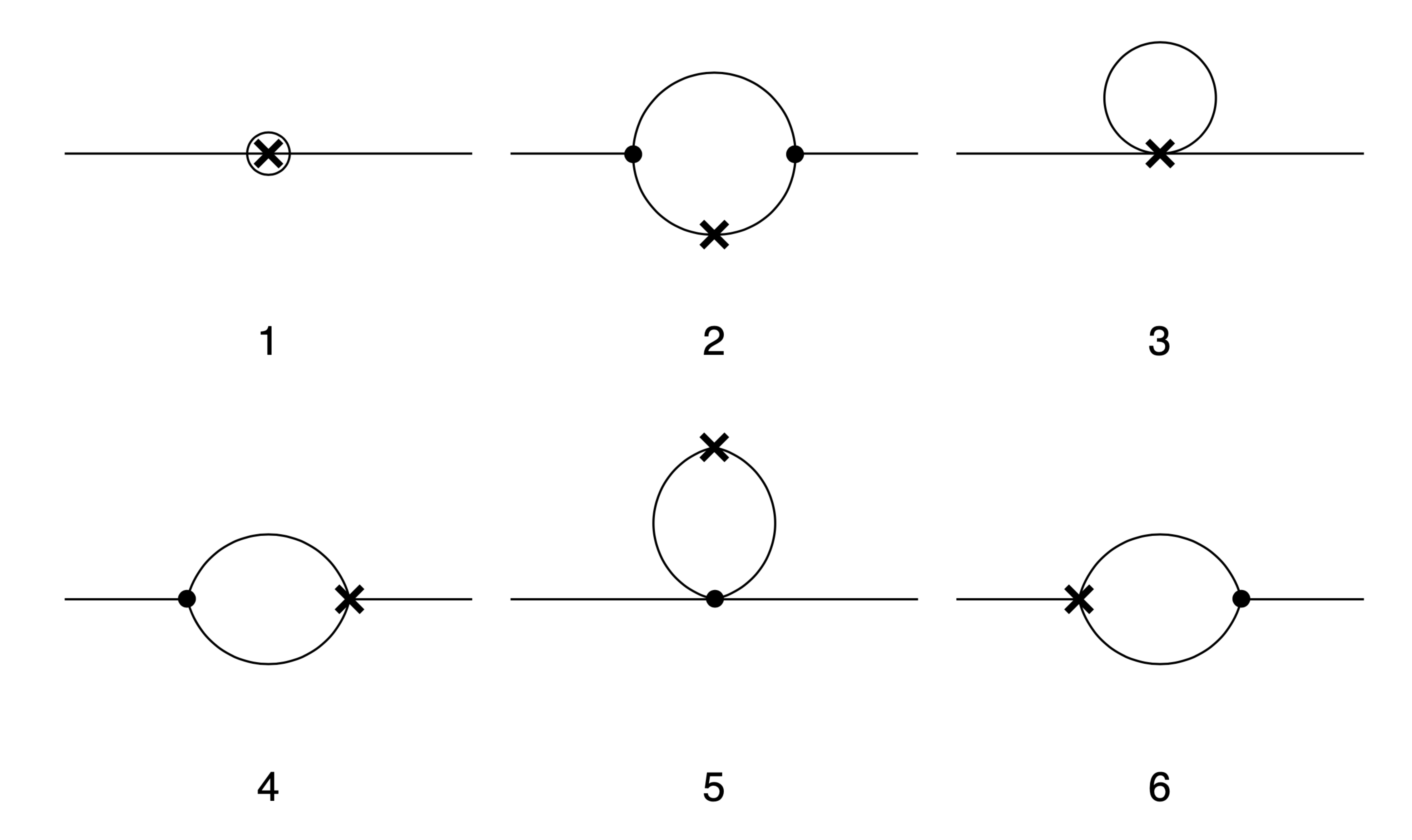}
	\caption{Topologies of the diagrams of counterterms in the Einstein-scalar QED model.}	\label{fig04b}
\end{figure}

\end{document}